# The Readout Supervisor Firmware for Controlling the Upgraded LHCb Detector and Readout System

Federico Alessio, Paolo Durante, Guillaume Vouters

*Abstract*— In 2019, the LHCb experiment at CERN will undergo a major upgrade where its detector electronics and the entire readout system will be replaced. The goal is to read-out all events at the full LHC frequency of 40 MHz, reaching a total data rate of ~40 Tb/s. In this context, a new timing, trigger and readout control system has been developed: its main tasks are to distribute centrally the clock, the timing information and to synchronize all elements in the readout system: from the very last Front-End ASIC to the building of events to be stored. The heart of the timing and readout control system is a VHDL firmware core that is now finalized and currently in use in test-benches and test-beams by the upgraded sub-detectors, for their initial commissioning phase. Such firmware core is able to generate all necessary processes to keep the synchronization of all readout elements with the LHC bunch crossing as well as it is able to generate asynchronous commands for calibration or test purposes. It is also able to accommodate for specific recipes to handle varying running conditions, by using a generic and fully configurable approach. In this paper, the philosophy and the implementation behind such firmware are described in details, posing particular emphasis to the real-time logical processes that were developed in order to satisfy the requirements of synchronization and readout control of the upgraded LHCb detector.

## I. Introduction

THE LHCb experiment at CERN [1] is devoted to the search for New Physics by precisely measuring its effects in CP violation and rare decays. By applying an indirect approach, LHCb is able to probe effects which are strongly suppressed by the Standard Model, such as those mediated by loop diagrams and involving flavor changing neutral currents. In the proton-proton collision mode, the LHC is to a large extent a heavy flavor factory producing over 100,000 bb-pairs every second at the nominal LHCb design luminosity of $2 \times 10^{32}$ $cm^{-2}$ $s^{-1}$. Given that bb-pairs are predominantly produced in the forward or backward direction, the LHCb detector was designed as a forward spectrometer with the detector elements installed along the main LHC beam line, covering a pseudo-rapidity range of $2 < \eta < 5$ well complementing the other LHC detectors ranges.

LHCb proved excellent performance in terms of data taking [2] and detector performance over the period 2010-2017 accumulating about 7 $fb^{-1}$ of data and it is foreseen to accumulate another ~2 $fb^{-1}$ over the final year 2018. The high efficiency (> 90%) and real-time solutions implemented during the data taking make the possibility of increasing the physics yield very attractive for the LHCb experiment. In fact, the LHCb detector is limited by design in terms of data bandwidth - 1 MHz instead of the LHC bunch crossing frequency of 40 MHz - and physics yield for hadronic channels at the hardware trigger. Therefore, a Letter Of Intent [3], a Framework TDR [4] and a Trigger and Online TDR [5] document the plans for an upgraded detector which will enable LHCb to increase its physics yield in the decays with muons by a factor of 10, the yield for hadronic channels by a factor 20 and to collect about 50 $fb^{-1}$ at a leveled constant luminosity of up to $2 \times 10^{33}$ $cm^{-2}$ $s^{-1}$. This corresponds to ten times the current design luminosity and increased pileup of a factor 5. The upgrade is foreseen to actually take place in the years 2019-2020.

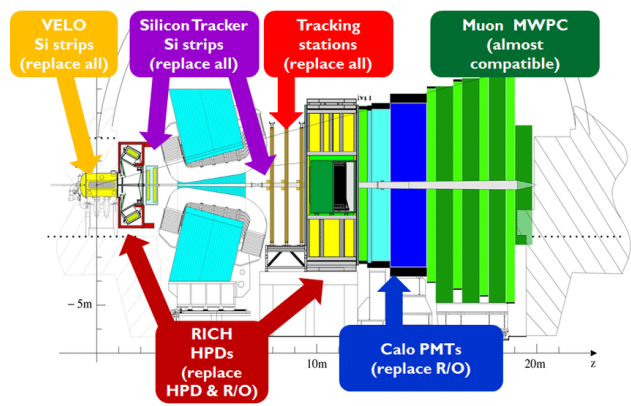

Fig. 1: The LHCb upgraded detector at CERN. The detector is built to perform precise vertexing, tracking, particle identification, calorimetry and muon detection. In the upgrade the entire Front-End and Back-End electronics will be changed to cope with a trigger-less readout and ~90% of detector channels will be changed as well. The detector will be majorly upgraded to increase its physics yields by a factor 10-20.

F. Alessio and P. Durante are with CERN, Rue de Meyrin, 1211 Geneva, Switzerland (e-mail: federico.alessio@cern.ch).
G. Vouters is with LAPP, BP 110, Annecy-le-Vieux, 74941 Annecy Cedex, France

## II. The Upgraded Readout architecture

In order to remove the main design limitations of the current LHCb detector, the strategy for the upgrade of the LHCb experiment essentially consists of ultimately removing the first-level hardware trigger entirely, hence to run the detector fully trigger-less. By removing the first-level hardware trigger, LHC events are recorded and transmitted from the Front-End electronics (FE) to the readout network at the full LHC bunch crossing rate of 40 MHz, resulting in a 40 Tb/s DAQ network. All events will therefore be available at the processing farm where a fully flexible software trigger will perform selection on events, with an overall output of about 20 kHz of events to disk.

Such approach will allow maximizing signal efficiencies at high event rates. The direct consequences of this approach are that some of the LHCb sub-detectors will need to be completely redesigned to cope with a higher average luminosity of $2\times10^{33}$ $cm^{-2}\ s^{-1}$ and the whole LHCb detector will be equipped with completely new trigger-less FE electronics (Fig. 1). In addition, the entire readout architecture must be redesigned in order to cope with the upgraded multi-Tb/s bandwidth and a full 40 MHz dataflow [6]. A view of the upgraded LHCb readout architecture can be seen in Fig. 2.

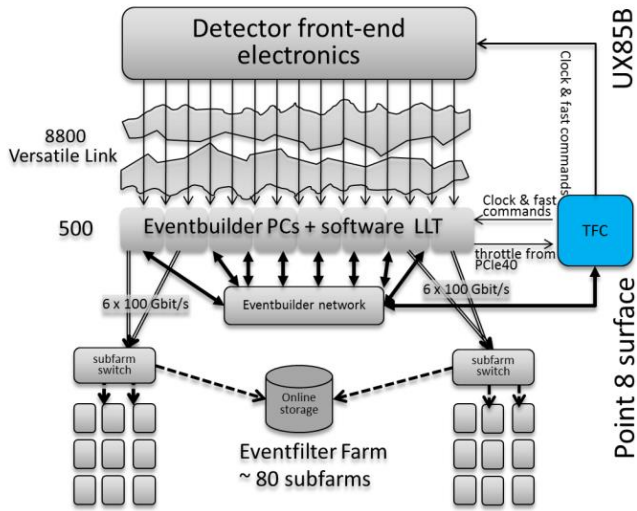

Fig. 2: The scheme of the upgraded LHCb readout architecture.

## III. The Upgraded LHCb Readout Control and Timing System

In this context, a new central system for timing distribution, readout control and event management for such upgraded readout system has been developed as well. It will replace the current readout control system, which will not be compatible in the scope of the upgrade of the LHCb experiment both in terms of hardware and in terms of specifications. In short, the upgraded readout control system is based on a single central Readout Supervisor that is responsible for:

- synchronizing and distributing the main bunch clock from the LHC, including its timing signals
- generating and distributing synchronous and asynchronous commands to all the Readout Boards and all the trigger-less FE, for a total of ~15000 destinations
- rate regulating the system and the rate at which events are sent to the farm by taking into account possible back-pressure from the event filter farm or by applying dedicated triggering/throttling logic to perform a central rejection of events based on predefined complex matrices of conditions.

More details about the specifications of the upgraded readout control system can be found in [7] and a scheme of the system can be seen in Fig. 3. The readout control system comprises:

- a single Readout Supervisor SODIN, in charge of centrally managing the readout architecture
- a set of Interface Boards (SOL40s) which will act as a fan-out for timing and readout control and as an interface board between the LHCb slow control and the FE electronics.

The communication with the various sub-systems in the architecture is ensured by a network of bidirectional optical links, up to ~2500 of them, and by usage of commercial FPGA technology and specific rad-hard ASICs. In addition, a custom-made PCIe card was developed in LHCb [8] to be the hardware backbone of the Data Acquisition system: the card is equipped with a powerful FPGA (Altera ARRIA 10), 48 bidirectional optical links qualified for up to 10 Gb/s, a 100 Gb/s PCIe Gen3 bus and a set of electrical interfaces. Such card will be widely used in LHCb and it will take the role of a Readout Board, should it be programmed with the Readout Board firmware, or the role of the Readout Supervisor, should it be programmed with the Readout Supervisor firmware. Each of these cards are coupled with a powerful server PC, for control of the FPGA and for interface to the upgraded readout system network.

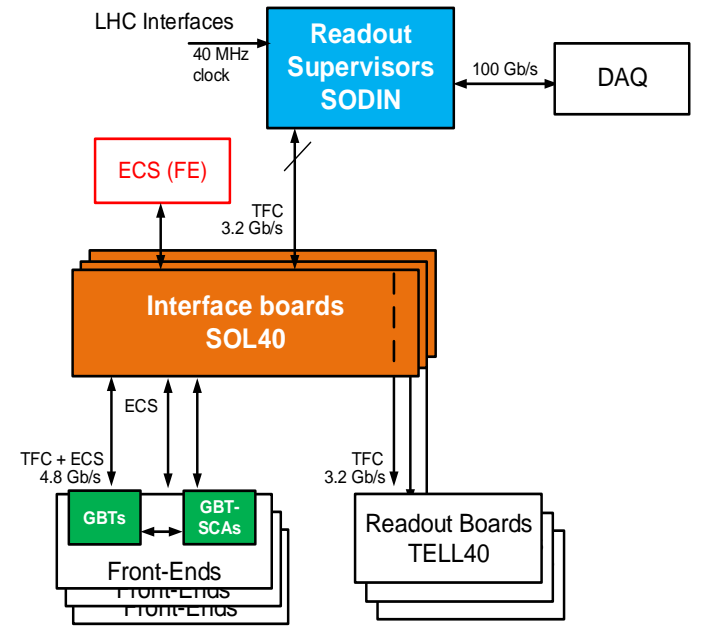

Fig. 3: The scheme of the upgraded LHCb readout control system.

Specifically, one of these PCIe cards will act as the single central Readout Supervisor of the final upgraded LHCb detector. It will be connected to a set of Interface Cards, which will have the role to fan-out the fast control information to the Front-End electronics and the Readout Boards, in order to reach all of the ~2500 destinations. A total of ~40 Interface Cards each equipped with 48 bidirectional links running at ~3.2 Gb/s are necessary to cover the full readout control network, including the use of optical splitters to the Readout Boards. Therefore, only a single FPGA will host the firmware code that contains the logic to satisfy the aforementioned specifications and to (fast) control the readout system. At the time of the abstract submission, such firmware has reached its first final version and it is currently widely used by every upgraded sub-detector in the LHCb experiment in test-beams and lab test-benches, for the pre-commissioning of the developed electronics and it is being fine-tuned for the final global commissioning of the upgraded detector, that is to start in 2019.

## IV. THE READOUT SUPERVISOR FIRMWARE

Since the beginning, the Readout Supervisor firmware for the upgraded LHCb detector was built and developed targeting already its final implementation, due to the very tight time constraint of the LHCb upgrade and in order to enforce global specifications of the full readout system, starting early on with the Front-End design phase. It has been done in a generic and programmable way, by using independent, parallel and generic processes, in order to satisfy the global requirements as specified above. At the same time, the firmware has been designed to satisfy specific extra-requirements that may arise during the pre-commissioning or commissioning of the upgraded sub-systems, should it be deemed necessary, without the need of having to produce different firmware versions at every ad-hoc request. It is also programmable and controllable at real-time, and it also handles external electrical input or software input from the control system. A simplified logical scheme of the firmware is shown in Fig. 4.

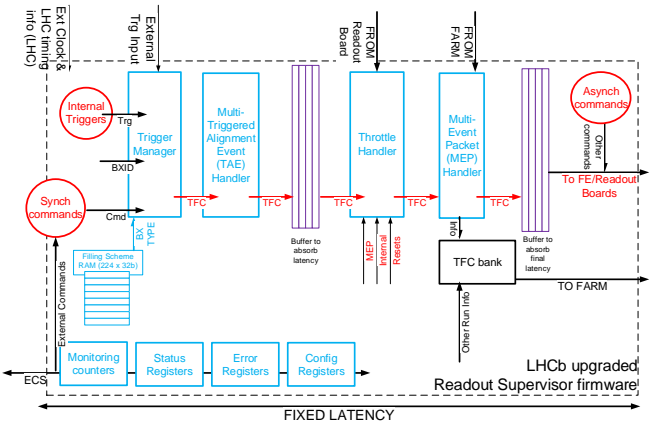

Fig. 4: Simplified logical scheme of the LHCb Upgraded Readout Supervisor firmware.

The main features of the Readout Supervisor firmware are individually described in the following sections.

### A. Synchronization with the global LHC clock.

The LHC clock is an electrical input to the LHCb PCIe card and it is the main clock of the system, used for the user logic and to clock the transceivers for optical distribution of the timing commands. In the PCIe card, the clock is first input to an external PLL (SI5345), cleaned and re-driven with the same exact LHC frequency and then transmitted to the FPGA as an input.

In addition, an LHC orbit signal is also received – through the same array of electrical inputs – and such orbit pulse is the signal used to reset the Bunch ID counter to identify events. Based on this signal, the firmware generates a *Bunch ID counter reset signal* which is part of the array of commands distributed to the entire readout electronics. In this way, every element in the system can reset their local Bunch ID counter based on an offset corresponding to the local delay. The Bunch ID is the unique identifier of an event and it is used in the Back-End (BE) to ultimately pack a full event from the individual fragments sent from the FE.

### B. Triggers generation

Even though the FE electronics runs fully trigger-less, a centralized trigger is generated in case rate regulation is needed (i.e. reduce the amount of accepted events). The trigger is not based on physics decisions, but on a predefined set of recipes and instructions loaded at real-time by the central control system and it is only distributed to the BE without any constraints on latency – although once the first trigger is generated, all subsequent triggers arrive at a fixed latency due to the synchronous nature of the timing and readout control system.

While in the final system, the triggering mechanism will be mostly unused, such technique is already playing a major role in the commissioning phase of the detector: to illustrate the concept with an example, the central firmware can in fact generate a calibration command and choose whether the data association to that calibration command should be accepted as an event or not. Should it be accepted, a trigger bit is associated to it to communicate to the BE boards to actually pack the fragments and send them out to the rest of the Data Acquisition System rather than rejecting the event. Also, for test-benches where there is no possibility of storing 40 MHz of events for many hours, a much lower trigger rate must be used.

The generation of a trigger can be associated with:
- internal processes like internal periodic trigger generators with specific configuration (like rate and location of the trigger)
- external electrical inputs, especially useful for test-beam or test-benches
- specific data taking modes
- individual asynchronous triggers triggered by the slow control

The firmware also implements a *throttle mechanism* that is the possibility of rejecting a possible accepted trigger due to a specific data taking reason. An example of this can be seen in the FE Reset procedure: when the central firmware generates an asynchronous FE Reset command for the entire FE electronics, it must also wait that the FE electronics is ready to transmit proper data again – making sure that the FE Reset procedure completed. In this sense, all triggers following the generation of the FE Reset and before a programmable waiting time are rejected, hence throttled away. This is only one example of the procedures, but such technique is widely used throughout the firmware in order to keep a centralized synchronicity with the full readout architecture.

*C. Generation of synchronous and asynchronous commands*

Probably the most important aspect in the Readout Supervisor firmware is the generation of centralized synchronous and asynchronous commands. Such commands are used by the FE and the BE electronics to perform specific actions or trigger specific processes, based on the global specifications.

The generation of synchronous commands is based on programmable processes and they can be triggered internally in the firmware – from another process for example – or externally from an electrical input for example. The generation of asynchronous commands is instead triggered via slow control (see section F).

An example of the way in which commands are generated, managed and then distributed can be seen in Fig. 5. This example describes a "start-of-run" synchronization mechanism. Such mechanism is used to synchronize the data stream of the FE to the input decoding block of the BE, such that the BE synchronize on the correct BXID for each individual link. This mechanism is done centrally in the Readout Supervisor and distributed to every individual FE chip (~2500) such that they perform a "synchronization mechanism" with the corresponding link in the BE card. The mechanism follows these steps:

1- the ECS sends a start of run command to the Readout Supervisor
2- the Readout Supervisor edge detects such command and generate a FE Reset and a BE Reset. These resets are one clock signal long.
3- the Readout Supervisor subsequently starts sending a command (Header Only) whose aim is to keep the FE idle. The length of the transmission of this command is programmable.
4- when the first Bunch Identifier Reset arrives after the first FE Reset, the Readout Supervisor releases the Header Only command and sends a Synch (synchronization) command, to have the FE start sending a synchronization pattern with the corresponding Bunch Identifier on which the BE will synchronize
5- the length of the Synch command is programmable and it may or may not be followed by another Header Only should some FE chip need to remain idle during the procedure.
6- once the procedure is done, all FE links are now synchronized with the BE and the Readout Supervisor can release the veto on the trigger, i.e. data can be accepted.

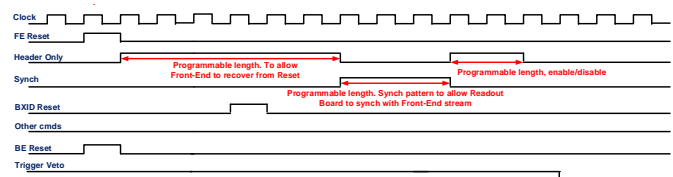

Fig. 5: Example of a start-of-run synchronization mechanism highlighting the interplay between synchronous and asynchronous commands, clock and specific processes.

A procedure of this kind can be repeated every time the control system sends a start of run command or even if it sends a FE Reset command, even asynchronously.

*D. Usage of programmable pipelines to ensure fixed latency transmission of commands*

In order to allow for variable and real-time triggering/throttle mechanisms due to varying running conditions while keeping the transmission of commands to the FE at a fixed latency, a set of programmable pipelines is used. In Fig. 4, two major pipelines are highlighted: the output pipeline for final latency adjustment and a middle pipeline which is used to generically compensate for specific data taking modes. An example of such specific data taking modes is the Timing Alignment Event (TAE) mode: in this mode – mostly used for calibration and commissioning purposes – a window of triggers and/or commands are generated around a central trigger in order to extend the window of the accepted events. In this case, triggers *"in the past"* must be modified in order to allow for the full window to be modified in the same way and this can easily be done with a fixed latency pipeline and modify the signals at the input of the pipeline. The current firmware allows for a full window up to 65 events.

*E. Generation of a centralized event data bank*

The generation of a centralized data bank containing the unique information of a selected event, such as timestamp, event type and sources and destination was deemed to be necessary and it is implemented in the firmware. As the Readout Supervisor is the only element in the system to centrally decide whether an event should be kept or not, it is also the only element in the system able to know the origin and the source of such event. Such information are:
- Bunch Identifier (BXID)
- Crossing Type based on the LHC filling scheme (beam-beam, beam-gas, empty-empty crossing types)
- Origin of trigger (calibration, TAE, random, etc.)
- A trigger mask that identifies if another command was associated to the accepted event (for example if the Front-End was requested to send data Non-Zero Suppressed – NZS).
- In case of a scan, the number of the step in the scan
- In addition, the bank is used to timestamp the event by uniquely identifying the event with an Orbit number – i.e. the number of orbit pulses since the beginning of a run – and the UTC timestamp as the number of 40 MHz clock cycles since the beginning of a run.

*F. Real-time interface to the central LHCb control system*

An important aspect of the central Readout Supervision is the ability to control and monitor the readout system in a fully real-time fashion. This is done partly by firmware processes as described in the previous sections, but it is also done by the software control part. In fact, the software control part is in charge of configuring the firmware, passing the most important parameters (such as programmable trigger rate, the masking of such triggers, the generation of commands, the initial timestamp, etc.) but also to generate asynchronous commands: for example, as described in section C, at the start of a run, an asynchronous FE Reset is generated and following such reset a start-of-run synchronization mechanism is done in firmware. However, the FE Reset is issued from control system, immediately after the start of run command and such synchronization mechanism will not take place until the central software system decides to send to. This is only one example, given here to indicate the way the interplay between control system and the firmware is done. It also highlights the high-level of control that the system can apply on the central readout modes: it can be chosen to be fast and synchronous for certain processes or routines, while it can be chosen to be asynchronous and slower for other routines, in case the start of run synchronization has to wait for another task to be finished before issuing a start of run.

The software control part is implement within the global LHCb Experiment Control System (ECS) framework [9].

*G. Monitoring registers and counters*

An important aspect of the central Readout Supervision is also the ability to monitor all the previously described processes. A choice was made to have a set of counters which are free-running and a set of counters which are "latched" to a specific update: such update is issued by control system and all counters are updated at the exact same clock cycle, such that the system can get a uniform and consistent update of the counters. For example, in this way, by comparing the number of calibration triggers and the number of orbits, as they are updated on the same clock cycles, the ratio of the two can be used to determine a frequency, very precisely and consistently.

*H. Amount of FPGA logical resources*

The firmware previously described has been fully implemented for an Altera Arria 10 FPGA and it is currently taking up to 3% of the Combinatorial ALUTs (tot of 8870) and 2% of the logic registers (7710) as well as 4% if the available memory bits (~70 kB).

The low amount of logical resources and memory resources are needed to satisfy the requirements in order to keep the FPGA routing simple and protected against timing issues as well as allowing to have many of such cores running in parallel in the same FPGA for partitioning purposes.

## V. CONCLUSION

In this paper, the firmware for the central Readout Supervisor of the LHCb upgraded readout system was presented. This firmware is currently being used for sub-detector commissioning, for test-beams and test-benches.

It is done in a generic, flexible and programmable way and in this paper the mechanisms behind the way synchronous and asynchronous commands, timing and procedures are generated is highlighted. This is also shown in relation to the fact that such commands should be distributed at a fixed latency and in relation to the interplay between the firmware and the software slow control.

Such firmware will be in production for the commissioning phase of the upgraded LHCb detector starting from the end of 2018 when the first sub-detector will start their commissioning and assembly and it will see its very final version for the first data taking at the LHC few years afterwards.

## REFERENCES

[1] The LHCb Collaboration, "The LHCb Detector at the LHC", 2008 *JINST* 3 S08005.
[2] F. Alessio on behalf of the LHCb Collaboration, "A true real-time success story: the case of collection beauty-ful data at the LHCb experiment", *this conference*.
[3] The LHCb Collaboration, "Letter of Intent for the LHCb Upgrade", *CERN-LHCC-2011-001,* 2011.
[4] The LHCb Collaboration, "Framework TDR for the LHCb Upgrade", *CERN-LHCC-2012-007,* 2012.
[5] The LHCb Collaboration, "LHCb Trigger and Online Upgrade Technical Design Report", *CERN-LHCC-2014-016,* 2014.
[6] F. Alessio et al., "Trigger-less readout architecture for the upgrade of the LHCb experiment at CERN", 2013 *JINST* 8 C12019
[7] F. Alessio and R. Jacobsson, "A new readout control system for the LHCb upgrade at CERN", 2012 *JINST* 7 C11010
[8] JP Cachemiche et al, "The PCIe-based readout system for the LHCb experiment", 2016 *JINST* 11 P02013
[9] C. Gaspar et al., "The LHCb experiment control system: on the path to full automation", *13th International Conference on Accelerator and Large Experimental Physics Control Systems (ICALEPCS2011).* Joint Accelerator Conference Website, 2011, pp. 20-23.